\documentclass{raa}
\usepackage{amsfonts}
\usepackage{amssymb}
\usepackage{amsmath}
\usepackage{booktabs}
\usepackage{graphicx}
\usepackage{lscape}
\usepackage{natbib}
\usepackage{mathrsfs}
\usepackage{mathbbol}
\usepackage{multirow}
\usepackage{textcomp}
\usepackage{txfonts}
\usepackage{longtable}
\usepackage{float}
\usepackage{ulem}

\bibpunct{(}{)}{;}{a}{}{,}

\begin{document}
\def\etal{et al.}
\def\Mgv{M_{\rm g,500}}
\def\Mg{M_{\rm gas}}
\def\YX {Y_{\rm X}}
\def\YSZ {Y_{\rm SZ}}
\def\YR {Y_{\rm 5R500}}
\def\Te {T_{\rm e}}
\def\Ne {n_{\rm e}}
\def\Mv {M_{\rm 500}}
\def\Rv {R_{500}}
\def\keV {\rm keV}
\def\YSZYX {Y_{\rm SZ}\--Y_{\rm X}}
\def\Plck {\emph{Planck}}
\def\xmm {\emph{XMM-Newton}}
\def\xn {\emph{XMM-Newton}}
\def\DA {D_{\rm A}}
\def\kB {k_{\rm B}}
\def\sigmaT {\sigma_{\rm T}}
\def\mec {m_{\rm e}c^2}
\def\Pe {P_{\rm e}}
\def\YSZC {Y_{\rm SZ,CMB}}
\def\YSZX {Y_{\rm SZ,X-ray}}
\def\YCX {\mbox{$Y_{\rm SZ,CMB}\--Y_{\rm SZ,X-ray}$}}
\def\YP {Y_{\rm SZ,\emph{Planck}}}
\def\YXMM {Y_{\rm SZ,\emph{XMM}}}
\def\YPYXMM {\mbox{$Y_{\rm SZ,\emph{Planck}}\--Y_{\rm SZ, \emph{XMM}}$}}
\def\PdX {\mbox{$Y_{\rm SZ,\emph{Planck}}/Y_{\rm SZ, \emph{XMM}}$}}
\def\PdYX {\mbox{$Y_{\rm SZ,\emph{Planck}}/Y_{\rm X}$}}
\def\YPYX {\mbox{$Y_{\rm SZ,\emph{Planck}}\--Y_{\rm X}$}}
\def\betamodel {\beta \ \rm model}
\def\Yv {Y_{500}}
\def\Dc {D_{\rm c}}

\title{The $\rm{Y_{SZ,Planck}\--Y_{SZ,XMM}}$ scaling relation and its difference between cool-core and non-cool-core clusters}

   \volnopage{Vol.0 (2019) No.0, 000--000}      
   \setcounter{page}{1}          

   \author{Y. Zhu\inst{1}
    \and Y.H. Wang\inst{1}
    \and H.H. Zhao\inst{2}
    \and S.M. Jia\inst{1}
    \and C.K. Li\inst{1}
    \and Y. Chen\inst{1}}

   \institute{
   Key Laboratory for Particle Astrophysics, Institute of High Energy Physics,
 Chinese Academy of Sciences, 100049 Beijing, China; {\it ychen@ihep.ac.cn}\\
        \and
            Department of Physics and Institute of Theoretical Physics,
 Nanjing Normal University, 210023 Nanjing, China \\
\vs\no
   {\small Received~~2018 September 20; accepted~~2019~February 17}}

  \abstract{We construct a sample of 70 clusters using data from \emph{XMM-Newton} and \emph{Planck} to investigate the \mbox{$Y_{\rm SZ,\emph{Planck}}\--Y_{\rm SZ, \emph{XMM}}$} scaling relation and the cool-core influences on the relation. $Y_{\rm SZ,\emph{XMM}}$ is calculated by accurate de-projected temperature and electron number density profiles derived from \emph{XMM-Newton}. $Y_{\rm SZ,\emph{Planck}}$ is the latest \emph{Planck} data restricted to our precise X-ray size $\theta_{\rm 500}$. To study the cool-core influences on \mbox{$Y_{\rm SZ,\emph{Planck}}\--Y_{\rm SZ, \emph{XMM}}$} scaling relation, we apply two criteria, limits of central cooling time and classic mass deposition rate, to distinguish cool-core clusters (CCCs) from non-cool-core clusters (NCCCs). We also use $Y_{\rm SZ,\emph{Planck}}$ from other papers, which are derived from different methods, to confirm our results.
   The intercept and slope of the \mbox{$Y_{\rm SZ,\emph{Planck}}\--Y_{\rm SZ, \emph{XMM}}$} scaling relation are
$A=-0.86 \pm 0.30$, $B=0.83 \pm 0.06$. The intrinsic scatter is $\sigma_{\rm ins}= 0.14 \pm 0.03$. The ratio of \mbox{$Y_{\rm SZ,\emph{Planck}}/Y_{\rm SZ, \emph{XMM}}$} is $1.03 \pm 0.05$, which is perfectly agreed with unity. Discrepancies of \mbox{$Y_{\rm SZ,\emph{Planck}}\--Y_{\rm SZ, \emph{XMM}}$} scaling relation between CCCs and NCCCs are found in observation. They are independent of cool-core classification criteria and $Y_{\rm SZ,\emph{Planck}}$ calculation methods, although discrepancies are more significant under the classification criteria of classic mass deposition rate. The intrinsic scatter of CCCs (0.04) is quite small compared to that of NCCCs (0.27). The ratio of \mbox{$Y_{\rm SZ,\emph{Planck}}/Y_{\rm SZ, \emph{XMM}}$} for CCCs is $0.89 \pm 0.05$, suggesting that CCCs' $Y_{\rm SZ,\emph{XMM}}$ may overestimate SZ signal. By contrast, the ratio of \mbox{$Y_{\rm SZ,\emph{Planck}}/Y_{\rm SZ, \emph{XMM}}$} for NCCCs is $1.14 \pm 0.12$, which indicates that NCCCs' $Y_{\rm SZ,\emph{XMM}}$ may underestimate SZ signal.
}
   \titlerunning{Y scaling relation}
   \authorrunning{Y. ZHU et al.}

   \maketitle

%
\section{Introduction}
Galaxy clusters are the largest gravitationally bound systems in the universe, formed by collapsing of matter under their self-gravity and merging of small clusters \citep{col1999, kra2012}. The process of formation is sensitive to the evolution of the universe, therefore the study of galaxy clusters can trace the growth of large-scale structure and constrain cosmological parameters \citep{sel2006,Vikhlinin2009iii,man2010i,roz2010,ben2013,planck2013xxix,planck2015xxiv}. Cluster mass is the most important quantity when using clusters as cosmological probes. However, directly measuring cluster mass is difficult because about 87\% of cluster mass is in the form of dark matter. Instead, we infer cluster mass through scaling relations with quantities that are convenient to observe, such as X-ray luminosity and temperature, velocity dispersion and thermal Sunyaev-Zel'dovich effect flux 
\citep{Arnaud2005, Maughan2007, Reichert2011, Zhang2011, Bohringer2013, Bocquet2015, Munari2013}. 

The thermal Sunyaev-Zel'dovich effect \citep[tSZ;][]{sun1980} describes a distortion of cosmic microwave background (CMB) spectrum caused by inverse Compton scattering of CMB photons off hot gas in intracluster medium (ICM). The integrated Compton parameter $\YSZ$ is obtained by the integration of tSZ signal over the cluster extent $V$, with the temperature $\Te$, and electron number density $\Ne$, as
\begin{equation}
\YSZ = \DA^{-2} \frac{\kB \sigmaT}{\mec} \int \Ne \Te dV = \DA^{-2} \frac{\sigmaT}{\mec} \int \Pe dV
\end{equation}
where $\Pe$ is the gas pressure, $\Pe = \Ne \kB \Te$ , $\kB$ is the Boltzmann constant, $\sigmaT$ is the Thomson cross section , $\mec$ is the electron rest mass and $\DA$ is the angular diameter distance. \cite{kra2006} introduce the $\YSZ$'s X-ray analogue, $\YX$, which is the product of cluster X-ray temperature $T_{\rm X}$ and gas mass $M_{\rm gas}$. Both $\YSZ$ and $\YX$ represent the total thermal energy of the cluster, therefore they are good mass proxies with low intrinsic scatter and with little relevance of the complicated dynamic state in clusters \citep{mot2005,nag2006,arn2007,zha2013,mah2013,sem2014iii}. We should note that to obtain the precise mass from the scaling relations, biases induced by the selection effects should be taken into account \citep{Pratt2009, all2011, Angulo2012, and2011}. $\YSZ$ has already been applied to derive cluster mass in some works, and they give serious consideration to possible bias to the mass proxy \citep{Aghanim2009, Comis2011, planck2011xi, Jimeno2018}.

$\YSZ$ can be obtained by two methods: 1) direct SZ observation, $\YSZC$; 2) SZ signal based on ICM properties derived from X-ray observation, $\YSZX$. $\YSZC$ is proportional to $\Ne \Te$ and relies more on the region outside the cluster core, while $\YSZX$ is sensitive to clumping regions because X-ray flux given by bremsstrahlung emission is proportional to $\Ne^2\Te^{1/2}$.  The different dependence of SZ and X-ray observations on $\Ne$ and $\Te$ may have influences on $\YSZC$-$\YSZX$ relation due to various physical processes in clusters. Therefore, the comparison between $\YSZC$ and $\YSZX$ may reveal discrepancies between cool-core clusters (CCCs) and non-cool core clusters (NCCCs), increasing knowledge of bias and intrinsic scatter of the SZ/X-ray scaling relation. Furthermore, unlike X-ray observation, SZ observation is not affected by surface brightness dimming, thus it is an ideal probe for galaxy clusters at high redshift. The SZ/X-ray scaling relation can be used to infer cluster mass, producing completive cosmology measurements.

Most previous works have focused on the relation between $\YSZ$ and $\YX$. Normally $\YSZC$ is not distinguished from $\YSZX$. They study $\YSZX$-$\YX$ scaling relation \citep{arn2010} and $\YSZC$-$\YX$ scaling relation, and find that the two relation are consistent with each other \citep{and2011, planck2011iii, roz2014i, roz2014ii, bif2014ii, Czakon2015}. Several papers study the $\YCX$ scaling relation \citep{bon2012,dem2016}, they also find good agreement between SZ signal and its X-ray prediction. Additionally, the outskirt of NCCCs has rich substructures, while that of CCCs is more homogeneous and relaxed. However, no discrepancy has been found between CCCs and NCCCs in observations so far \citep{planck2011viii,roz2012,dem2016}.

In the following, we use a sample of 70 clusters to determine the $\YCX$ scaling relation. Accurate ICM properties, derived from $\xn$ data analyzed with the $\beta$ model and the de-projected method, are applied to calculate $\YSZX$. On the other hand, $\YSZC$ is obtained from the $\Plck$ latest catalogue. Every quantity in our analysis, e.g. $\Te$, $\Ne$, is directly from observations, independent of assumed scaling relations which are widely used in other works to infer some quantities. This would reduce artificial correlations introduced in data processing, and improve the reliability of our results.

The paper is organized as follows. In Section 2 we introduce the cluster sample and describe the SZ and X-ray data analysis, respectively. In Section 3 we investigate the $\YCX$ scaling relation and the influences of CCCs and NCCCs on this relation. The discussions about our results are also presented. We make a conclusion in Section 4.

We assume a flat $\Lambda$CDM cosmology with $\Omega_{\rm M}= 0.3$, $\Omega_{\rm \Lambda}= 0.7$, and $H_0 = 70 \rm{km/s/Mpc}$. All uncertainties are quoted at the 68\% confidence level.

\section{Data}

\subsection{Cluster Sample}
Our sample is extracted from $\xn$ bright cluster sample (XBCS) \citep{zhathesis, zha2015} and Planck PSZ2 catalogue \citep{planck2015xxvii}. For the XBCS, we select the clusters with a flux-limited ($ f_X[0.1-2.4keV] \geq 1.0\times 10^{-11} ~\rm erg~s^{-1}~cm^{-2}$) method from several cluster catalogues based on \emph{ROSAT} All-Sky Survey \citep[RASS;][]{deg1999}: HIghest X-ray FLUx Galaxy Cluster Sample \citep[HIFLUGCS;][]{rei2002}, \emph{ROSAT}-ESO Flux Limited X-ray catalogue \citep[REFLEX;][]{boh2004}, Northern \emph{ROSAT} Brightest Cluster Sample \citep[NORAS;][]{boh2000i}, X-ray-bright Abell-type clusters \citep[XBACs][]{ebe1996i}, \emph{ROSAT} Brightest Cluster Sample \citep[BCS;][]{ebe1998i}. Among the XBCS, 78 clusters are available in PSZ2 catalogue. 
The position of cluster center identified by $\xn$ and $\Plck$ has some deviation. Clusters with conditions of $\Delta D > 4'$ or $\Delta D>0.3 \Rv$ are excluded, where $\Delta D$ is the positional offset between two centers, $\Rv$ is the cluster radius where the mean density is that 500 times the critical density of the universe at the cluster redshift. Our final sample consists of 70 clusters, covering the redshift from about 0.01 to 0.25. The mass within $R_{500}$ of these galaxy clusters ranges from 0.27 to 11.5 $\rm 10^{14}M_{\odot}$, while the $R_{500}$ ranges from 0.44 to 2.45 $\rm Mpc$.

\subsection{$\Plck$ data}
PSZ2 catalogue is constructed by the blind detection over full sky using three independent extraction algorithms: MMF1, MMF3, PsW, with no prior positional information on known clusters. MMF1 and MMF3 are based on matched-multi-frequency filter algorithm. PsW is a fast Bayesian multi-frequency algorithm. All the three algorithms assume the generalized Navarro-Frenk-White (GNFW) pressure profile \citep{arn2010} as the cluster prior spatial characteristics, given by
\begin{equation}
p(r)=\frac {P_0}{(c_{500}r / \Rv)^{\gamma} [1+(c_{500}r /\Rv)^{\alpha}]^{(\beta-\gamma) / \alpha}}
\end{equation}
with the parameters \citep{planck2013xxix}
\\$[P_0,c_{500},\gamma,\alpha,\beta]=[8.40h_{70}^{-3/2}, 1.177,0.308,1.05,5.49]$,
\\where $\alpha$, $\beta$, $\gamma$ are the logarithmic slopes for the intermediate region ($c_{500}r \sim \Rv$), the outer region ($c_{500}r \gg \Rv$) and the core region ($c_{500}r \ll \Rv$), respectively, $c_{500}$ is the concentration parameter through which $\theta_{500}$ (instead of radial coordinates, angular coordinates are more often used, as $\theta_{500} = \Rv / \DA$) is related to the characteristic cluster scale $\theta_{\rm s}$ ($\theta_{\rm s}=\theta_{500} / c_{500}$), and $P_0$ is the normalization factor. $\theta_{\rm s}$ and $P_0$ are free parameters in this profile.

For each detected source, each algorithm provides an estimated position, $S/N$ value, a two-dimensional joint probability distribution for $\theta_{\rm s}$ and the integrated Compton parameter within $5\theta_{\rm 500}$, $\YR$ \citep[see][fig.16]{planck2015xxiv}.

$\YR$ and $\theta_{\rm s}$ are strongly correlated, we adopt $\theta_{\rm 500}$, equivalently $\theta_{\rm s}$, which is accurately derived from $\xn$ observation (see 2.3) to break the $\YR - \theta_{\rm s}$ degeneracy. Given the $\theta_{\rm s}$ from X-ray, the expectation and standard deviation from the $\YR$ conditional distribution are derived as the value of $\YR$ and its uncertainty. Uncertainties less than $0.05\YR$ would be assigned to the standard deviation of $\YR$ in PSZ2, because they may be slightly underestimated by such $\YR$ estimation \citep{planck2015xxvii}. Finally, $Y_{500}$, denoted as $\YP$, is converted from $\YR$ by $\YR = 1.79 \cdot Y_{500}$ for the pressure profile mentioned above \citep{arn2010,planck2013xxix}.

\subsection{XMM-Newton data}
The XBCS is built using a flux-limited method, We elaborately process the XMM-Newton data of the whole cluster sample.
Here only a brief description of the $\xn$ data process is presented, and more details can be referred to \cite{zha2013,zha2015}. $\xn$ pn/EPIC data acquired in Extended Full Frame mode or Full Frame mode are performed with Science Analysis System (SAS) 12.6.0. We select events with FLAG=0 and PATTERN$\leq$4, in which contaminated time intervals are discarded. Then we correct vignetting effects and out-of-time events, remove prominent background flares and point sources, and subtract the particle background and the cosmic X-ray background. After that the cluster region is divided into several rings centered on the X-ray emission peak, with the width of the rings depending on the net photon counts. 
{Point Spread Function \citep[PSF, pn: FWHM = 6";][]{Handbook2018} effect can be ignored because the minimum width of rings is set at 30".
By subtracting all the contributions from outer regions, the de-projected spectrum of each ring is obtained \citep{chen2003,chen2007,jia2004,jia2006}.

XSPEC version 12.8.1 is used for spectral analysis. De-projected temperature $\Te$, metallicity and normalizing constant $norm$ at each ring are derived from the de-projected spectra fits with the thermal plasma emission model Mekal \citep{mew1985} and Wabs model \citep{mor1983}. Fitting the simulated spectrum using $\Te$ and abundance profiles in XSPEC, one can get the de-projected electron number density $\Ne$ at each ring.

Limited by $\xn$ field of view and the statistics of photons from clusters, the maximum observable radius of clusters, $R_{\rm max}$, is usually smaller than $\Rv$. In the case of $R_{\rm max} < \Rv$, $\Te$ at $r > R_{\rm max}$ is set to the same value in the outermost ring. Linear interpolation is used to calculate $\Te (r)$. For the fits of electron density profile $\Ne (r)$, single $\betamodel$ and double $\betamodel$ are both  adopted.

The single $\betamodel$ gives
\begin{equation}
\Ne (r) = n_0 [1+ (\frac {r}{r_{\rm c}})^2]^{- \frac {3}{2} \beta}
\end{equation}
where $n_0$ is the electron number density, and $r_{\rm c}$ is the core radius.

Double $\betamodel$ is in the form of
\begin{equation}
\Ne (r) = n_{01} [1+ (\frac {r}{r_{\rm c1}})^2]^{- \frac {3}{2} \beta_1} + n_{02} [1+ (\frac {r}{r_{\rm c2}})^2]^{- \frac {3}{2} \beta_2}
\end{equation}
where $n_{01}$ and $n_{02}$ are the electron number density, $r_{c1}$ and $r_{c2}$ are the core radius for the inner and outer components \citep{chen2003}.

For most clusters, double $\betamodel$ fits better than single $\betamodel$ significantly, however for some clusters, the improvements are neglectable. As a result, 54 and 16 clusters are fitted with double and single $\betamodel$, respectively. Fig.\ref{fig:A0576} shows a typical cluster profile. It clearly indicates that double $\betamodel$ matches the electron number density data better than single $\betamodel$.

The influences of the center offsets $\Delta D$ between $\xn$ observation and three $\Plck$ algorithm detections are considered. Because of the $\Plck$ blind detection, we cannot fix our X-ray cluster position to the $\Plck$ detection procedure and re-extract $\YP$. Instead we correct the $\YXMM$ by changing its integral center.
The cluster is assumed to be spherically symmetric, $\YXMM$ within $\Rv$ is given by

\begin{equation}
\Yv = \DA^{-2} ~ \frac{\kB \sigmaT}{\mec} ~\int\limits^{\Rv+\Delta D}_{-\Rv+\Delta D}
~~\int\limits^{R_y}_{-R_y} ~~\int\limits^{R_z}_{-R_z}
\Ne \Te dxdydz
\end{equation}
where $R_y = \sqrt{\Rv^2-x^2}$ and $R_z = \sqrt{\Rv^2-x^2-y^2}$.

We adopt the Monte-Carlo method to estimate the uncertainties of $\YXMM$. For each cluster, we simulate $\Te$ at each shell, and the parameters of $\betamodel$ for $\Ne$ profile 5000 times, following Gaussian distributions with their own uncertainties. Then the uncertainty of $\YXMM$ is obtained.

\begin{figure}[H]
\centering
  \includegraphics{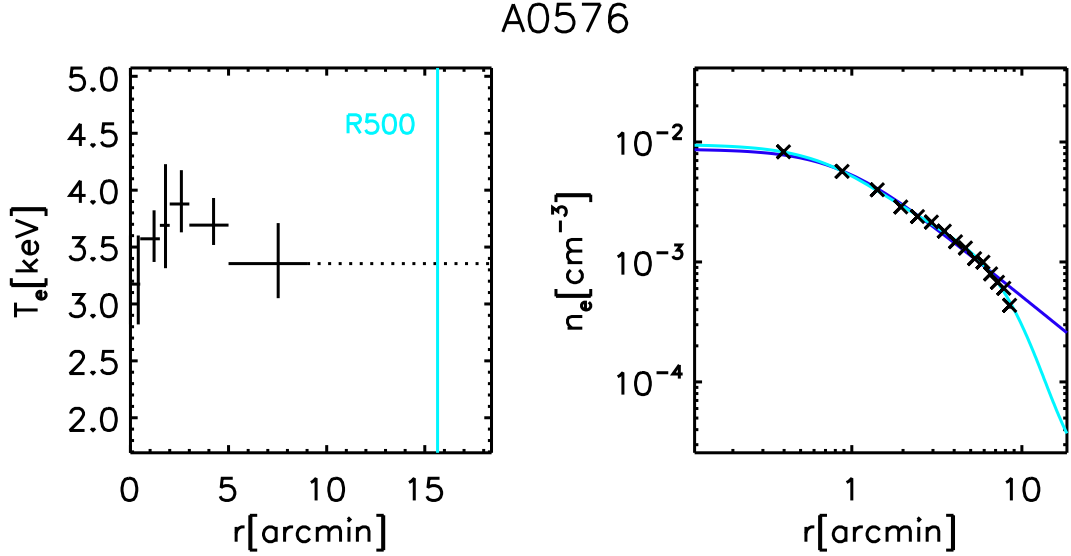}
      \caption{Illustration of profiles for cluster A0576. \emph{Left panel}: temperature (marked as cross symbol) with error bar at each ring. Light blue vertical line indicates the position of $\Rv$. Extrapolation of temperature is shown as dotted line. \emph{Right panel}: electron number density (marked as cross symbol) at each ring. Light blue line and deep blue line indicate the density profiles fitting by single $\betamodel$ and double $\betamodel$, respectively.}
\label{fig:A0576}
\end{figure}

\section{Results and discussion}
\subsection{Fitting method}
Emcee is the affine-invariant ensemble sampler for Markov chain Monte Carlo (MCMC) designed for Bayesian parameter estimation \citep[][the code can be downloaded in http://dan.iel.fm/emcee/current/]{for2013}. We employ emcee to fit the $\YCX$ scaling relation in the linear form
\begin{equation}
Y=B \cdot X + A
\end{equation}
where $A$ and $B$ are estimated parameters, $X$ and $Y$ denote the base-10 logarithm of $\YSZX$ and $\YSZC$ ($\log_{10}\YSZX$, $\log_{10}\YSZC$), respectively. Likelihood adopted in these fits is from the equation (35) of \cite{hog2010}, following \cite{planck2015xxvii},
\begin{equation}
\ln L =  - \frac{1}{2} \sum_{i=1}^{N}{\Big{(}\ln (\sigma_{\rm i}^2 + \sigma_{\rm int}^2)+ \sum_{i=1}^{N}{\frac {(Y_{\rm i} - B \cdot X_{\rm i} -A)^2}{\sigma_{\rm i}^2 + \sigma_{\rm int}^2}}\Big{)}}
\end{equation}
where $\sigma_{\rm i}^2 = \sigma_{Y_{\rm i}}^2 + B^2 \cdot \sigma_{X_{\rm i}}^2$. $N$ is the number of clusters, $\sigma_{\rm int}$ is the intrinsic scatter, $\sigma_{X_{\rm i}}$ and $\sigma_{Y_{\rm i}}$ are statistical errors of the $X_{\rm i}$ and $Y_{\rm i}$. Three parameters $A$, $B$ and $\sigma_{\rm int}$ are estimated in the fitting procedure.
We also fix $B=1$, and repeat the procedure above to obtain $A$ and $\sigma_{\rm ins \verb"|" B=1}$. The ratio of $\YSZC/\YSZX$ equals to $10^A$.

\subsection{$\YP$ versus $\YXMM$}
$\YP$ and $\YXMM$ are all integrated within $\Rv$. We construct five samples named as MMF1, MMF3, PsW, MaxSN and NEAREST. $\YP$ in MMF1, MMF3, PsW samples are given by the three corresponding $\Plck$ extraction algorithms with fixing $\theta_s$ in ($\YR$,$\theta_{\rm s}$) probability distribution plane at the X-ray $\theta_s$. Four clusters are discarded from PsW sample because the X-ray $\theta_s$ is beyond the scope of the PsW ($\YR$,$\theta_{\rm s}$) plane. \cite{planck2015xxvii} proved that the detections characteristics made by three algorithms are consistent with each other by simulation. In order to construct a larger sample, we make use of them to build the MaxSN and NEAREST samples. In MaxSN sample, the $\YP$ of each cluster is assigned by the algorithm which gives the maximum $S/N$ (signal to noise) value, while in NEAREST sample, the $\YP$ of each cluster is set by the algorithm whose output position is closest to the X-ray center. With the accurate de-projected temperature and density distributions, we calculate $\YXMM$ correcting the impacts of the center offsets between $\xn$ and three $\Plck$ algorithms. Cluster properties are listed in Table \ref{tab:properties}, differences between $\YXMM$ in sample MaxSN and that in NEAREST are less than 2\%, therefore we only present $\YXMM$ in NEAREST sample in this table.

The scaling relations between $\YP$ and $\YXMM$ are shown in Fig.~\ref{fig:Y_all}. The best-fitting parameters and the number of clusters for each sample are presented in Table \ref{tab:all}.
 Firstly, we compare MMF1, MMF3 and PsW Samples which are constructed by three independent detection algorithms. On the condition that the slope and normalization are free parameters, $\YPYXMM$ relation in these three samples agree with each other. The intrinsic scatter in MMF1 sample is relatively larger than that in other algorithms. When we consider the relation with slope fixed to 1 ($B=1$), the ratio of $\PdX$ of MMF1 sample is significantly higher ($\sim4\sigma$) than that of MMF3 and PsW samples.
This is due to the different background estimations and extraction strategies in the different algorithms.
 For the combined samples, MaxSN and NEAREST, $\YPYXMM$ relation between them are consistent. We regard the NEAREST sample as our reference sample, because detection significance in each algorithm is different between blind mode and the mode with a prior known cluster position, and detection which provide the position closest to the cluster's X-ray center is considered to be the most accurate detection.

 NEAREST sample contains 70 clusters, in which 18, 18, 34 detections are respectively made by algorithm MMF1, MMF3, PsW, confirming that PsW produces the most accurate positions \citep{planck2015xxvii}. The intercept and slope of the $\YPYXMM$ relation in this sample are $A=-0.86 \pm 0.30$, $B=0.83 \pm 0.06$. The intrinsic scatter is $\sigma_{\rm ins} =0.14 \pm 0.03$. The ratio of $\PdX$ is $1.03 \pm 0.05$ which perfectly agrees with unity. Our results indicate that the SZ signal detected by CMB and by X-ray observation are fully consistent.

There are two papers that study on the $\YCX$ scaling relation. \cite{bon2012} present a sample of 25 massive relaxed galaxy clusters observed by Sunyaev Zel'dovich Array (SZA) and \emph{Chandra}. They assume the ICM model introduced by \cite{bul2010}  which can be applied simultaneously to SZ and X-ray data. Their ratio of $Y_{\rm SZ,CMB}/Y_{\rm SZ,X-ray}$ is $1.06 \pm 0.04$, which is in good agreement with our results. \cite{dem2016} use a sample of 560 clusters whose properties are derived from $\Plck$ 2013 foreground cleaned Nominal maps and \emph{ROSAT} observations, to determine SZ/X-ray scaling relations.

They calculate the angular size weighted $Y_{\rm SZ}$, obtain the relation $\bar Y_{\rm{SZ,} \Plck}= 0.97 \bar Y_{\rm SZ,X-ray}$, which also agrees with ours.


\begin{figure}
\centering
  \includegraphics[width=\textwidth, keepaspectratio]{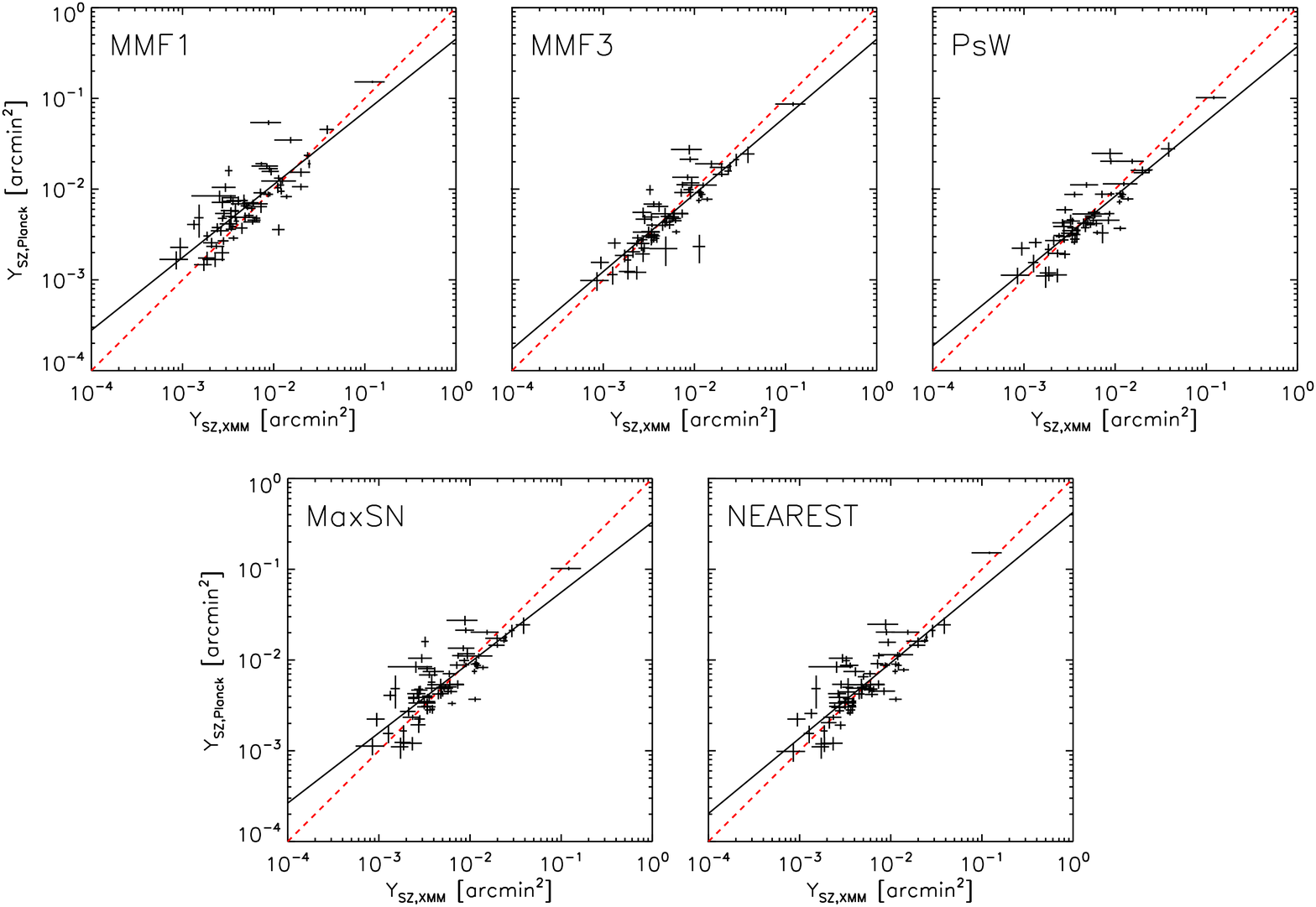}
      \caption{Scaling relations between $\YXMM$ and $\YP$. $\YXMM$ is modified by the cluster center differences between X-ray and the algorithms used to determine $\YP$. \emph{Top panels}: $\YP$ is measured using MMF1, MMF3 and PsW algorithm, respectively. \emph{Bottom panels}: combination of the three algorithms. \emph{Bottom left}: $\YP$ is determined by the most significant detection algorithm. \emph{Bottom right}: $\YP$ is assigned by the algorithm which gives closest position from X-ray center. The solid black lines represent the best fit lines, and the dashed red lines show the relations of $X=Y$.}
      \label{fig:Y_all}
\end{figure}

   \begin{table}
     \caption[]{$\YPYXMM$ scaling relations of five samples.}
     \label{tab:all}
  \centering
    \begin{tabular}{lcccccc}
\hline\hline
    Sample & N & A & B & $\sigma_{\rm ins}$ & $\PdX$* & $\sigma_{\rm ins \verb"|" B=1}$ \\
 \hline
    $\rm{MMF1}$ &	$67$	& $-0.79 \pm 0.36$ &	$0.80 \pm 0.07$ &	$0.20 \pm 0.04$	 & $1.27 \pm 0.08$ & $0.22 \pm 0.05$   \\
    $\rm{MMF3}$ & $66$ & $-0.80 \pm 0.26$ & $0.85 \pm 0.05$ & $0.10 \pm 0.03$ & $0.95 \pm 0.05$ & $0.11 \pm 0.03$ \\
    $\rm{PsW}$ & $61$ & $-0.99 \pm 0.28$ & $0.82 \pm 0.05$ & $0.11 \pm 0.03$ & $0.93 \pm 0.05$ & $0.13 \pm 0.03$ \\
    \hline
    $\rm{MaxSN}$ & $^a 70$ & $-1.11 \pm 0.31$ & $0.77 \pm 0.06$ & $0.17 \pm 0.04$ & $1.07 \pm 0.06$ & $0.20 \pm 0.04$ \\
    $\rm{NEAREST}$ & $^b 70$ & $-0.86 \pm 0.30$ & $0.83 \pm 0.06$ & $0.14 \pm 0.03$ & $1.03 \pm 0.05$ & $0.15 \pm 0.04$ \\
 \hline
  \end{tabular}
\tablecomments{0.86\textwidth}{The cluster number contributed by each algorithm to MaxSN and NEAREST samples:
$^{a}$ MMF1: 16, MMF3: 29, PsW: 25; $^{b}$ MMF1: 18, MMF3: 18, PsW: 34.\\
 * $\PdX = 10^{A|B=1}$. \\}
\end{table}

The intrinsic scatter in our results $\sigma_{\rm ins} = 0.14 \pm 0.03$ is slightly larger than the prediction ($\sim 10\%$). The extrapolation in both $\Plck$ and $\xn$ may induce scatter or bias to our results. When determining $\YP$, $\Yv$ is obtained from $Y_{\rm 5R500}$. The shape of the GNFW pressure profile employed in $\Plck$ analysis is fixed, which leaves neglectable impact to the scaling relation \citep{planck2011xi}, but different shapes of pressure profile may have significantly different conversion factors from $Y_{\rm 5R500}$ to $\Yv$ \citep{say2016}. To be more specific, each cluster has a unique pressure profile and a unique conversion factor, converting $\Yv$ from $Y_{\rm 5R500}$ by a unified factor may induce scatter. In the extrapolation of X-ray's cluster properties, a flat temperature extended from $\sim 0.5\Rv$ to the cluster's outer region could overestimate $\YXMM$.

We also calculate the $\YXMM$ whose $\Ne(r)$ fitting only with the single $\betamodel$, the ratio is $\PdX =0.89 \pm 0.05$, nearly 3$\sigma$ deviated from our previous result. Many studies argue that the isothermal $\betamodel$ is inadequate to fit ICM and may overestimate the SZ signal \citep{lie2006,bie2007,hal2007,Atrio-Barandela2008,mro2009,Allison2011}. Assuming two components in ICM in fitting electron distribution, double $\betamodel$ works well within $\Rv$ \citep{chen2007}.


\subsection{Cool core influences}
We construct a subsample including 55 clusters which are overlapping clusters between the HIFLUGCS and ours. In this subsample, we refer data in NEAREST sample to investigate the cool core influences on the scaling relations.
We adopt two methods to distinguish CCCs from NCCCs using X-ray data. The first method follows the definition in \cite{zha2013} (hereafter Z13): clusters with the central cooling time $t_{\rm c}<7.7h_{70}^{-1/2}$ (Rafferty et al. 2006) and the temperature drop larger than 30\% from the peak are classified as CCCs, it divides the sample into 28 NCCCs and 27 CCCs. The second method follows the definition in \cite{chen2007} (hereafter C07): clusters with significant classical mass deposition rate $\dot{M} \geq 0.01M_{\odot}yr^{-1} $ are classified as CCCs. Instead of calculating the mass deposition rate by ourselves, we directly use its classification which divides the sample into 29 NCCCs and 26 CCCs.

Fig.~\ref{fig:clstYSZcc} shows the CCCs' and NCCCs' scaling relations between $\YP$ and $\YXMM$. The best-fit parameters for each subsample are presented in Table~\ref{tab:clstYSZcc}.

In Z13 classification criteria, intrinsic scatter of $\YPYXMM$ scaling relation of CCCs ($\sim 0.11$) is slightly smaller than that of NCCCs ($\sim 0.20$), and the $\PdX$ ratio of CCCs trends to be less than that of NCCCs. Due to the relatively large uncertainties, we observe weak evidence for the discrepancies between CCCs and NCCCs. Under C07 criteria, disagreements between CCCs and NCCCs become more significant, especially for the intrinsic scatter which is $\sim 0.04$ and $\sim 0.28$ of CCCs and NCCCs, respectively. These results are not only obtained in NEAREST sample, they remain the same in other samples, which are shown in Table~\ref{tab:nrst_cc_all}.

To validate our results, we use $\YP$ taken from three papers, $Y_{500}$ in \cite{planck2011xi}, $Y_{\rm z}$ in PSZ1 catalogue \citep{planck2013xxix} and $Y_{\rm blind}$ in PSZ2 catalogue \citep{planck2015xxvii}, to discuss the cool-core influences on $\YPYXMM$ scaling relations. $Y_{500}$ in \cite{planck2011xi} is obtained by algorithm re-extraction from $\Plck$ maps at X-ray position and with the X-ray size. $Y_{\rm z}$ in PSZ1 is calculated using redshift information. $Y_{\rm blind}$ in PSZ2 is the blind detection which is bias high on average because of over-estimated size. Our $\YPYXMM$ is derived from $Y_{\rm blind}$ restricting with our X-ray size. Under C07 cool-core criteria, CCCs and NCCCs show clearly discrepancies on the SZ and X-ray measurements no matter which $\YP$ we used. Results are listed in Table~\ref{tab:ypyxmm_psz} and shown in Fig.~\ref{fig:ypyxmm_psz}.
\begin{figure}
   \centering
 \includegraphics[width=0.99\linewidth]{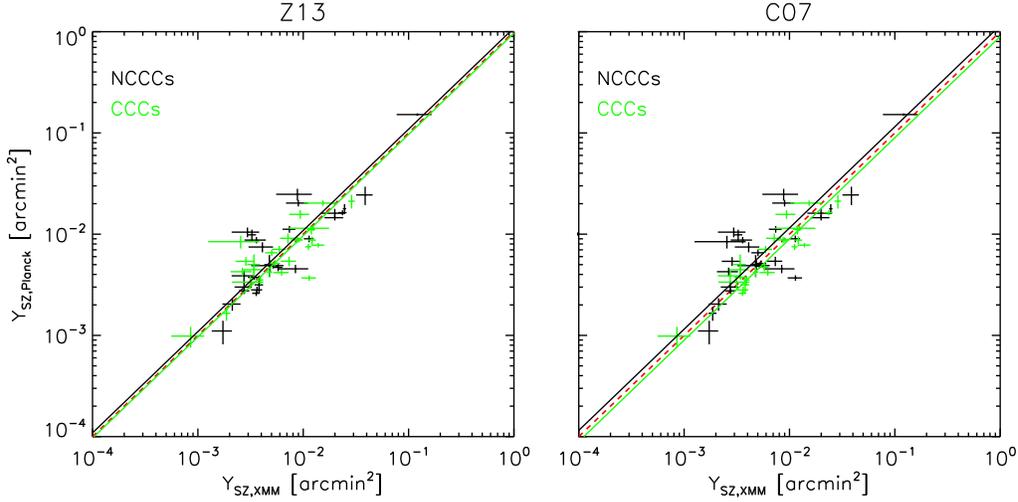}
   \caption{$\YPYXMM$ scaling relations for CCCs and NCCCs in NEAREST subsample. Z13 criteria \citep[left,][]{zha2013} and C07 criteria \citep[right,][]{chen2007} on CCCs and NCCCs are shown. Black dots indicate NCCCs, and green dots indicate CCCs. The black and green solid lines are the best fit lines of $\PdX$ for NCCCs and CCCs, respectively. The dashed red lines show the relations of $X=Y$.}
\label{fig:clstYSZcc}%

\end{figure}
\begin{table}
\caption{$\YPYXMM$ scaling relations for CCCs and NCCCs in NEAREST subsample with two cool core classification criteria.}
\label{tab:clstYSZcc}
\centering
\begin{tabular}{cccccccc}     
\hline\hline
 Class. & Sample & N & A & B & $\sigma_{\rm ins}$ & $\PdX$ & $\sigma_{\rm ins \verb"|" B=1}$ \\
 \hline
 \multirow{2}{1cm}{$^a$ Z13} & NCCCs &28 & $-0.54 \pm 0.49$ & $0.88 \pm 0.10$ &	$0.20 \pm 0.07$ & $1.09 \pm 0.10$ &	$0.20 \pm 0.07$\\
                    & CCCs  &27	& $-1.33 \pm 0.52$ & $0.74 \pm 0.10$ &	$0.11 \pm 0.04$ & $0.97 \pm 0.08$ & $0.13 \pm 0.05$\\
 \hline
 \multirow{2}{1cm}{$^b$ C07} & NCCCs &29	& $-0.84 \pm 0.57$ & $0.81 \pm 0.11$ &	$0.27 \pm 0.09$ & $1.14 \pm 0.12$ & $0.28 \pm 0.09$\\
	                & CCCs	&26	& $-0.78 \pm 0.36$ & $0.86 \pm 0.07$ &  $0.04 \pm 0.02$ & $0.89 \pm 0.05$ &	$0.04 \pm 0.02$\\
\hline
\end{tabular}
\tablecomments{0.86\textwidth}{$^{a}$ \cite{zha2013};~~ $^{b}$ \cite{chen2007}.\\}

\end{table}
\begin{table}
\caption{$\YPYXMM$ scaling relations for CCCs and NCCCs with two cool core classification criteria.}
\label{tab:nrst_cc_all}
\centering
\begin{tabular}{ccccccccc}     
\hline\hline
 Sample & Class. & Sub-Sample & N & A & B & $\sigma_{\rm ins}$ & $\PdX$ & $\sigma_{\rm ins \verb"|" B=1}$ \\
 \hline
\multirow{4}{1.5cm}{MMF1}	&	\multirow{2}{1cm}{Z13}	&	NCCCs &  28  &  $-0.32 \pm 0.54$ &  $0.87 \pm 0.11$&  $0.28 \pm 0.10$ &  $1.40 \pm 0.15$ & $0.28 \pm 0.09$ \\
	&		&	CCCs &  25 &   $-1.71 \pm 0.59$ & $0.64 \pm 0.12$ &   $0.12 \pm 0.05$ &  $1.14 \pm 0.10$ & $0.16 \pm 0.06$ \\
\cline{2-9}	
&	\multirow{2}{1cm}{C07}	&	NCCCs &  29 &    $-0.62 \pm 0.61$ & $0.81 \pm 0.12$ & $0.34 \pm 0.11$ &  $1.40 \pm 0.17$ & $0.35 \pm 0.11$\\
	&		&	CCCs &  24 &  $-0.93 \pm 0.52$ &  $0.79 \pm 0.10$  &  $0.08 \pm 0.04$ &  $1.13 \pm 0.08$ & $0.09 \pm 0.04$\\
\hline
\multirow{4}{1.5cm}{MMF3}	&	\multirow{2}{1cm}{Z13}	&	NCCCs &  26 &  $-0.70 \pm 0.46$ &  $0.84 \pm 0.09$ &  $0.17 \pm 0.06$ &  $1.08 \pm 0.10$ & $0.18 \pm 0.07$\\	
	&		&	CCCs &  26  &  $-1.00 \pm 0.40$ &  $0.83 \pm 0.08$ &  $0.05 \pm 0.03$ &  $0.87 \pm 0.05$ & $0.05 \pm 0.03$\\
\cline{2-9}		
&	\multirow{2}{1cm}{C07}	&	NCCCs &  27 &  $-0.72 \pm 0.51$ &  $0.85 \pm 0.10$ &  $0.22 \pm 0.09$ &  $1.03 \pm 0.10$ & $0.22 \pm 0.08$\\
	&		&	CCCs &  25  &  $-0.97 \pm 0.36$ &  $0.83 \pm 0.07$ &  $0.04 \pm 0.02$ &  $0.88 \pm 0.05$ & $0.05 \pm 0.02$\\
\hline
\multirow{4}{1.5cm}{PsW}	&	\multirow{2}{1cm}{Z13}	&	NCCCs &  24  &  $-0.42 \pm 0.45$ &  $0.92 \pm 0.09$ &  $0.12 \pm 0.06$ &  $0.99 \pm 0.08$ & $0.12 \pm 0.06$\\
	&		&	CCCs &  23 &  $-1.87 \pm 0.52$ &  $0.66 \pm 0.10$ &  $0.09 \pm 0.04$ &  $0.85 \pm 0.07$ & $0.14 \pm 0.06$\\
\cline{2-9}		
&	\multirow{2}{1cm}{C07}	&	NCCCs &  24 &  $-0.83 \pm 0.58$  &  $0.83 \pm 0.11$ &  $0.25 \pm 0.09$ &  $1.01 \pm 0.11$ & $0.25 \pm 0.10$\\
	&		&	CCCs &  23 &  $0.82 \pm 0.07$  &  $0.02 \pm 0.01$ &  $-1.07 \pm 0.34$ &  $0.82 \pm 0.04$ & $0.03 \pm 0.02$\\
\hline
\hline
\multirow{4}{1.5cm}{MaxSN}	&	\multirow{2}{1cm}{Z13}	&	NCCCs &  28  &  $-0.87 \pm 0.49$ &  $0.79 \pm 0.10$ &  $0.23 \pm 0.08$ &  $1.20 \pm 0.13$ & $0.26 \pm 0.09$\\
	&		&	CCCs &  27 &    $-1.42 \pm 0.52$ & $0.73 \pm 0.10$ &  $0.11 \pm 0.04$ &  $0.94 \pm 0.08$ & $0.13 \pm 0.05$\\
\cline{2-9}		
&	\multirow{2}{1cm}{C07}	&	NCCCs &  29  &  $-1.02 \pm 0.60$ &  $0.76 \pm 0.12$ &  $0.30 \pm 0.10$ &  $1.19 \pm 0.14$ & $0.34 \pm 0.11$\\
	&		&	CCCs &  26  &  $-1.12 \pm 0.39$ &  $0.79 \pm 0.08$ &  $0.05 \pm 0.02$ &  $0.94 \pm 0.06$ & $0.07 \pm 0.03$\\
\hline
\multirow{4}{1.5cm}{NEAREST}	&	\multirow{2}{1cm}{Z13}	&	NCCCs &  28  &  $-0.54 \pm 0.49$ &  $0.88 \pm 0.10$ &  $0.20 \pm 0.07$ &  $1.09 \pm 0.10$ & $0.20 \pm 0.07$\\
	&		&	CCCs &  27 &    $-1.33 \pm 0.52$ &$0.74 \pm 0.10$ &  $0.11 \pm 0.04$ &  $0.97 \pm 0.08$ & $0.13 \pm 0.05$\\
\cline{2-9}		
&	\multirow{2}{1cm}{C07}	&	NCCCs &  29  &  $-0.84 \pm 0.57$ &  $0.81 \pm 0.11$&  $0.27 \pm 0.09$ &  $1.14 \pm 0.12$ & $0.28 \pm 0.09$\\
	&		&	CCCs &  26 &   $-0.78 \pm 0.36$ &$0.86 \pm 0.07$ &   $0.04 \pm 0.02$ &  $0.89 \pm 0.05$ & $0.04 \pm 0.02$\\

\hline
\end{tabular}
\end{table}

\clearpage
\begin{figure}
  \centering
  \includegraphics[width=0.99\linewidth]{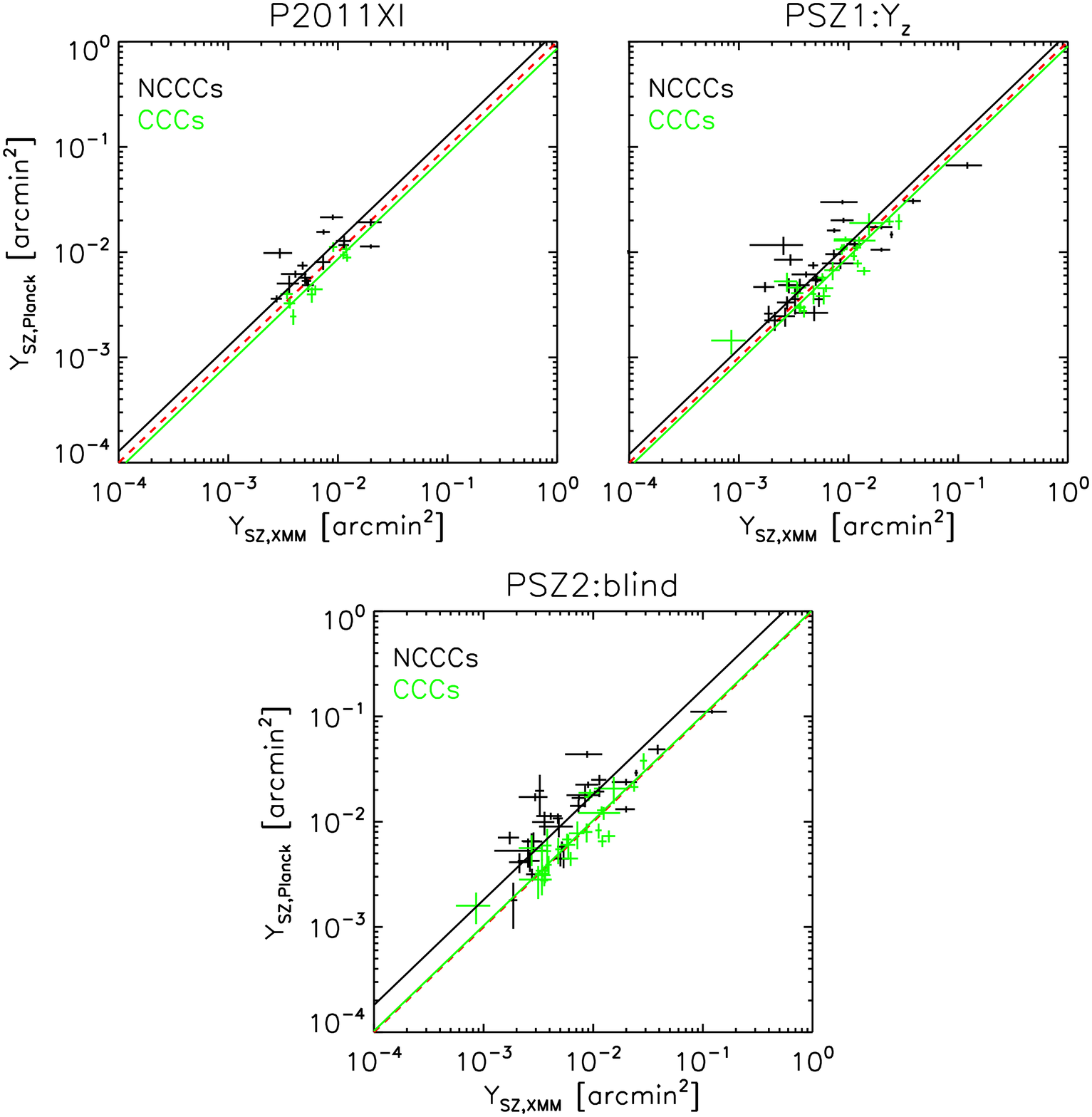}\\
  \caption{$\YPYXMM$ scaling relations for CCCs and NCCCs with different $\YP$ under C07 criterion.}
  \label{fig:ypyxmm_psz}
\end{figure}
\begin{table}[H]
\renewcommand{\arraystretch}{1.05}
\caption{$\YPYXMM$ scaling relations for CCCs and NCCCs with different $\YP$ under C07 criterion.}
\label{tab:ypyxmm_psz}
\centering
\begin{tabular}{ccccccccc}
\hline\hline
  Class. & $\YP$ & Sample & N & A & B & $\sigma_{\rm ins}$ & $\PdX$ & $\sigma_{\rm ins \verb"|" B=1}$ \\
 \hline
\multirow{8}{1cm}{C07} & \multirow{2}{1.8cm}{$^a$ p2011XI} & NCCCs &  15 &  $-1.40 \pm 0.92$ &  $0.67 \pm 0.18$ &  $0.13 \pm 0.07$ &  $1.28 \pm 0.14$ & $0.14 \pm 0.08$ \\
           &  & CCCs &  10 &  $-0.23 \pm 0.75$ &  $0.98 \pm 0.15$ &  $0.03 \pm 0.03$ &  $0.86 \pm 0.06$ & $0.02 \pm 0.03$ \\
 \cline{2-9}
 & \multirow{2}{1.8cm}{$^b$ PSZ1:$Y_{\rm z}$} & NCCCs &  29 &  $-1.06 \pm 0.45$ &  $0.75 \pm 0.09$ &  $0.15 \pm 0.06$ &  $1.20 \pm 0.11$ & $0.19 \pm 0.07$ \\
           &  & CCCs &  23 &  $-0.94 \pm 0.46$ &  $0.83 \pm 0.09$ &  $0.06 \pm 0.03$ &  $0.90 \pm 0.06$ & $0.06 \pm 0.03$  \\
 \cline{2-9}

  & \multirow{2}{1.8cm}{$^c$ PSZ2:blind} & NCCCs &  29 &  $-0.66 \pm 0.52$ &  $0.75 \pm 0.10$ &  $0.21 \pm 0.07$ &  $1.81 \pm 0.19$ & $0.25 \pm 0.09$ \\
           &  & CCCs &  26 &  $-0.54 \pm 0.57$ &  $0.89 \pm 0.11$ &  $0.11 \pm 0.04$ &  $1.03 \pm 0.08$ & $0.10 \pm 0.04$ \\
\hline
\end{tabular}
\tablecomments{0.86\textwidth}{$^{a}$ \cite{planck2011xi};~~ $^{b}$ \cite{planck2013xxix};~~$^{c}$ \cite{planck2015xxvii}\\}
\end{table}

We also study the $\YPYX$ scaling relation. Compared with $\YXMM$ which requires accurate temperature and electron number density distribution, $\YX$, which equals to mean temperature multiplied by gas mass, is much easier to obtain. Therefore the $\YPYX$ scaling relation is more widely used in comparing SZ and X-ray data. Here we define $\YX=T_{\rm X} \cdot M_{\rm gas} \cdot (\DA^{-2}(\sigmaT / \mec)/(\muup_{\rm e} m_{\rm p}))$  , where $T_{\rm X}$ is the volume average temperature determined within the region $[0.2,0.5]\Rv$, $M_{\rm gas}$ is the gas mass within $\Rv$, $4\pi m_{\rm p} \int_0^{\Rv} \Ne (r) r^2 dr$, with $m_{\rm p}$ the proton mass and $\muup_{\rm e}$ the mean molecular weight of the electrons,  the factor $\DA^{-2}(\sigmaT / \mec)/(\muup_{\rm e} m_{\rm p})$ is used to convert the unit from $\rm Mpc^2$ to $\rm arcmin^2$. 

$\YPYX$ relations, with C07 and Z13 criteria, 
are shown in Fig.~\ref{fig:clstYXcc}. 
We find similar results in the $\YPYX$ relation as in $\YPYXMM$ relation, which indicate that SZ and X-ray observations on CCCs and NCCCs are inconsistent, although discrepancies of  Y-ratio between CCCs and NCCCs in $\YPYX$ relation are smaller than that in $\YPYXMM$ relation, intrinsic scatters of CCCs and NCCCs still significantly disagree with each other. Results are listed in Table ~\ref{tab:clstYXcc}. We emphasize that $\PdYX = 0.92 \pm 0.05$ is completely consistent with the prediction in X-ray $0.924 \pm 0.004$ \citep{arn2010}.  

\begin{figure}
   \centering
 \includegraphics[width=0.99\linewidth]{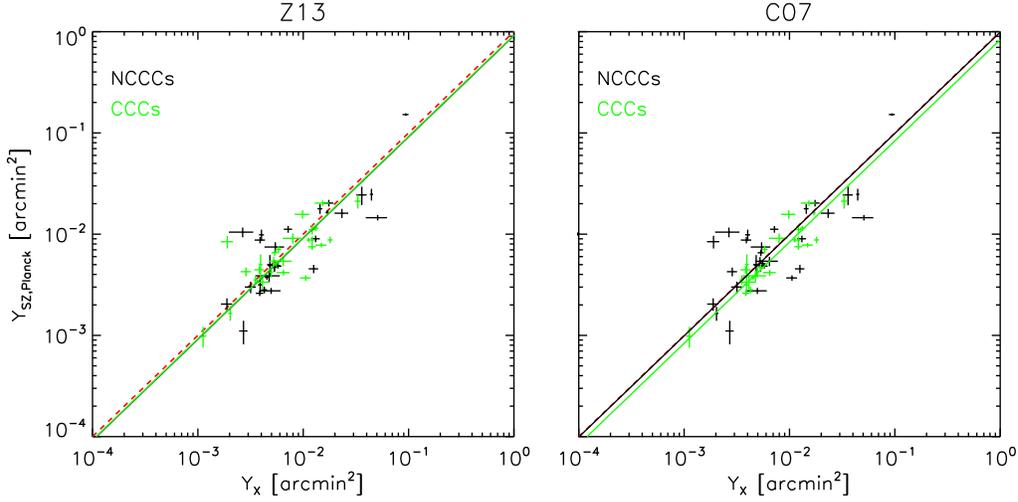}
   \caption{$\YPYX$ Scaling relations for CCCs and NCCCs in NEAREST subsample. The convention of lines and panels is the same as Fig.~\ref{fig:clstYSZcc}.}
   \label{fig:clstYXcc}%

\end{figure}
\begin{table}
\caption{$\YPYX$ scaling relations for CCCs and NCCCs in NEAREST subsample with two cool core classification criteria.}
\label{tab:clstYXcc}
\centering
\begin{tabular}{cccccccc}     
\hline\hline
 Class. & Sample & N & A & B & $\sigma_{\rm ins}$ & $\PdYX$ & $\sigma_{\rm ins \verb"|" B=1}$ \\
 \hline
 \multirow{2}{1cm}{Z13} & NCCCs &28 & $-0.75 \pm 0.54$ & $0.87 \pm 0.11$ & $0.29 \pm 0.09$ & $0.92 \pm 0.10$ & $0.29 \pm 0.09$\\
                    & CCCs  &27	& $	-1.62 \pm 0.55	$ & $	0.69 \pm 0.11	$ & $	0.16 \pm 0.05	$ & $	0.92 \pm 0.08	$ & $	0.20 \pm 0.06	$ \\
 \hline
 \multirow{2}{1cm}{C07} & NCCCs &29	& $	-1.27 \pm 0.58	$ & $	0.74 \pm 0.12	$ & $	0.36 \pm 0.11	$ & $	0.99 \pm 0.12	$ & $	0.41 \pm 0.12	$\\
	                & CCCs	&26	& $	-0.60 \pm 0.40	$ & $	0.91 \pm 0.08	$ & $	0.07 \pm 0.03	$ & $	0.84 \pm 0.05	$ & $	0.06 \pm 0.02	$ \\

\hline
\end{tabular}
\end{table}

Our sample is an intersection of the X-ray sample with flux limit, and the Planck sample with S/N cut. The selection effects of Malmquist bias \citep{Stanek2006} and Eddington bias \citep{Maughan2007} may deviate the results due to scatters in these scaling relations around limit/cut. To quantify these effects on scaling relations, complicated computations are required to generate large mock clusters sample from assumed mass function, to mimic the observed sample with the same selection criteria \citep{Vikhlinin2009, planck2011viii, planck2011xi, roz2012, Czakon2015, dem2016}. For Y-ratio, the correction is negligible according to \cite{planck2011xi, roz2012, Czakon2015, dem2016}. Bias should be fairly small for very lumious objects \citep{planck2011x, roz2012, planck2015xxvii}. As galaxy clusters in our sample are very bright clusters with strong SZ detections, we believe the bias of Eddington effect and Malmquist effect is fairly small in our $\YPYXMM$ scaling relation. The discrepencies between CCCs and NCCCs are due to other reasons. However, we should also bear in mind that our $\YPYXMM$ scaling relation is derived from most luminous clusters. Applications to dimmer clusters with this scaling relation should be careful.

Most CCCs are relaxed systems while NCCCs are undergoing more disturbing processes, like merging. Therefore the intrinsic scatter of CCCs is smaller than that of NCCCs. The ratio of $\PdX$ in CCCs (NCCCs) has a trend to be smaller (larger) than unity, which implies that the outskirt pressure profiles of CCCs and NCCCs could have substantial differences, instead of following a universal profile.

Because of the different dynamic state between CCCs and NCCCs, it's natural to believe that $\YCX$ scaling relation of CCCs and NCCCs could have discrepancies, but previous measurements show little difference between them \citep{planck2011v,roz2012,dem2016}. This contradiction may be mainly due to our high quality X-ray data. We detailedly process the $\xn$ data, and no scaling relation is referred during data analyzation. Another reason may be due to cool-core classification criteria. In our results, the CCCs and NCCCs discrepancies are more significant with the C07 definition, therefore the mass deposition rate may be much closer to the physical nature of CCCs and NCCCs than the central gas density, core entropy excess and central cooling time which previous works apply to distinguish CCCs from NCCCs.


\section{Conclusion}
In this paper we use a sample of 70 clusters to study the $\YPYXMM$ scaling relations and compare the differences between CCCs and NCCCs. The $\YXMM$ is calculated by accurate de-projected temperature and electron number density profiles derived from $\xn$, with correction of cluster center offset between two satellites, and the $\YP$ is the latest $\Plck$ data restricted to our X-ray cluster size $\theta_{\rm 500}$. We build five samples: MMF1, MMF3, PsW, MaxSN and NEAREST, while the MaxSN and NEAREST samples are the combinations of MMF1, MMF3, PsW. \\
The results in MaxSN and NEARESET samples are in fully agreement, and we choose NEAREST sample as our reference. The intercept and slope of the $\YPYXMM$ scaling relation are
$A=-0.86 \pm 0.30$, $B=0.83 \pm 0.06$. The intrinsic scatter is $\sigma_{\rm ins}= 0.14 \pm 0.03$. The ratio of $\PdX$ is $1.03 \pm 0.05$, which is perfectly agree with unity.\\
We use two classification criteria to distinguish CCCs from NCCCs. Both criteria indicate that the properties of CCCs are inconsistent with that of NCCCs. The intrinsic scatter of CCCs is significantly small compared with that of NCCCs, and the ratio of $\PdX$ of CCCs (NCCCs) has slight inclination to be smaller (larger) than unity, suggesting that $\YXMM$ for CCCs (NCCCs) may overestimate (underestimate) SZ signal. Discrepancies under criterion of C07 are more significant than that of Z13. We study $\YPYXMM$ relation using other $\YP$ taken from three $\Plck$ papers, and we also investigate $\YPYX$ relation in the same way. We find that cool-cores do have influences on SZ/X-ray scaling relations. Therefore we draw a firm conclusion that the intrinsic scatter and the $\PdX$ ratio of CCCs disagree with that of NCCCs.

\begin{acknowledgements}
      Thank Dr. Yang Yan-Ji and Dr. Fang Kun for helpful discussions.\\
      This research is supported by: Bureau of International Cooperation, Chinese Academy of Sciences (GJHZ1864). H.H. Zhao acknowledges support from the National Natural Science Foundation of China under grant 11703014.\\
 
\end{acknowledgements}

\bibliographystyle{raa}
\bibliography{Yscalingrelaiton_raa_vfinal}

\begin{thebibliography}{72}
\providecommand\natexlab[1]{#1}
\providecommand\JournalTitle[1]{#1}

\bibitem[{Aghanim} {et~al.}(2009)]{Aghanim2009}
{Aghanim}, N., {da Silva}, A.~C., \& {Nunes}, N.~J. 2009, \aap, 496, 637

\bibitem[{Allen} {et~al.}(2011)]{all2011}
{Allen}, S.~W., {Evrard}, A.~E., \& {Mantz}, A.~B. 2011, \araa, 49, 409

\bibitem[{Allison} {et~al.}(2011)]{Allison2011}
{Allison}, J.~R., {Taylor}, A.~C., {Jones}, M.~E., {Rawlings}, S., \& {Kay},
  S.~T. 2011, \mnras, 410, 341

\bibitem[{Andersson} {et~al.}(2011)]{and2011}
{Andersson}, K., {Benson}, B.~A., {Ade}, P.~A.~R., {et~al.} 2011, \apj, 738, 48

\bibitem[Angulo {et~al.}(2012)]{Angulo2012}
Angulo, R.~E., Springel, V., White, S.~D., {et~al.} 2012, \mnras, 426, 2046

\bibitem[Arnaud {et~al.}(2005)]{Arnaud2005}
Arnaud, M., Pointecouteau, E., \& Pratt, G.~W. 2005, \aap, 903, 893

\bibitem[{Arnaud} {et~al.}(2007)]{arn2007}
{Arnaud}, M., {Pointecouteau}, E., \& {Pratt}, G.~W. 2007, \aap, 474, L37

\bibitem[{Arnaud} {et~al.}(2010)]{arn2010}
{Arnaud}, M., {Pratt}, G.~W., {Piffaretti}, R., {et~al.} 2010, \aap, 517, A92

\bibitem[{Atrio-Barandela} {et~al.}(2008)]{Atrio-Barandela2008}
{Atrio-Barandela}, F., {Kashlinsky}, A., {Kocevski}, D., \& {Ebeling}, H. 2008,
  \apj, 675, L57

\bibitem[{Benson} {et~al.}(2013)]{ben2013}
{Benson}, B.~A., {de Haan}, T., {Dudley}, J.~P., {et~al.} 2013, \apj, 763, 147

\bibitem[{Bielby} \& {Shanks}(2007)]{bie2007}
{Bielby}, R.~M., \& {Shanks}, T. 2007, \mnras, 382, 1196

\bibitem[{Biffi} {et~al.}(2014)]{bif2014ii}
{Biffi}, V., {Sembolini}, F., {De Petris}, M., {et~al.} 2014, \mnras, 439, 588

\bibitem[{Bocquet} {et~al.}(2015)]{Bocquet2015}
{Bocquet}, S., {Saro}, A., {Mohr}, J.~J., {et~al.} 2015, \apj, 799, 214

\bibitem[B{\"{o}}hringer {et~al.}(2013)]{Bohringer2013}
B{\"{o}}hringer, H., Chon, G., Collins, C.~A., {et~al.} 2013, \aap, 555, A30

\bibitem[{B{\"o}hringer} {et~al.}(2000)]{boh2000i}
{B{\"o}hringer}, H., {Voges}, W., {Huchra}, J.~P., {et~al.} 2000, \apjs, 129,
  435

\bibitem[{B{\"o}hringer} {et~al.}(2004)]{boh2004}
{B{\"o}hringer}, H., {Schuecker}, P., {Guzzo}, L., {et~al.} 2004, \aap, 425,
  367

\bibitem[{Bonamente} {et~al.}(2012)]{bon2012}
{Bonamente}, M., {Hasler}, N., {Bulbul}, E., {et~al.} 2012, New Journal of
  Physics, 14, 025010

\bibitem[{Bulbul} {et~al.}(2010)]{bul2010}
{Bulbul}, G.~E., {Hasler}, N., {Bonamente}, M., \& {Joy}, M. 2010, \apj, 720,
  1038

\bibitem[{Chen} {et~al.}(2003)]{chen2003}
{Chen}, Y., {Ikebe}, Y., \& {B{\"o}hringer}, H. 2003, \aap, 407, 41

\bibitem[{Chen} {et~al.}(2007)]{chen2007}
{Chen}, Y., {Reiprich}, T.~H., {B{\"o}hringer}, H., {Ikebe}, Y., \& {Zhang},
  Y.-Y. 2007, \aap, 466, 805

\bibitem[{Colberg} {et~al.}(1999)]{col1999}
{Colberg}, J.~M., {White}, S.~D.~M., {Jenkins}, A., \& {Pearce}, F.~R. 1999,
  \mnras, 308, 593

\bibitem[Comis {et~al.}(2011)]{Comis2011}
Comis, B., {De Petris}, M., Conte, A., Lamagna, L., \& {De Gregori}, S. 2011,
  \mnras, 418, 1089

\bibitem[Czakon {et~al.}(2015)]{Czakon2015}
Czakon, N.~G., Sayers, J., Mantz, A., {et~al.} 2015, 806, arXiv:1406.2800

\bibitem[{De Grandi} {et~al.}(1999)]{deg1999}
{De Grandi}, S., {B{\"o}hringer}, H., {Guzzo}, L., {et~al.} 1999, \apj, 514,
  148

\bibitem[{De Martino} \& {Atrio-Barandela}(2016)]{dem2016}
{De Martino}, I., \& {Atrio-Barandela}, F. 2016, \mnras, 461, 3222

\bibitem[{Ebeling} {et~al.}(1998)]{ebe1998i}
{Ebeling}, H., {Edge}, A.~C., {Bohringer}, H., {et~al.} 1998, \mnras, 301, 881

\bibitem[{Ebeling} {et~al.}(1996)]{ebe1996i}
{Ebeling}, H., {Voges}, W., {Bohringer}, H., {et~al.} 1996, \mnras, 281, 799

\bibitem[{Foreman-Mackey} {et~al.}(2013)]{for2013}
{Foreman-Mackey}, D., {Hogg}, D.~W., {Lang}, D., \& {Goodman}, J. 2013, \pasp,
  125, 306

\bibitem[{Hallman} {et~al.}(2007)]{hal2007}
{Hallman}, E.~J., {Burns}, J.~O., {Motl}, P.~M., \& {Norman}, M.~L. 2007, \apj,
  665, 911

\bibitem[Handbook(2018)]{Handbook2018}
Handbook, X.-n.~U. 2018, {XMM-Newton}, Vol. 2018

\bibitem[{Hogg} {et~al.}(2010)]{hog2010}
{Hogg}, D.~W., {Bovy}, J., \& {Lang}, D. 2010, arXiv:1008.4686

\bibitem[{Jia} {et~al.}(2006)]{jia2006}
{Jia}, S.-M., {Chen}, Y., \& {Chen}, L. 2006, \cjaa, 6, 181

\bibitem[{Jia} {et~al.}(2004)]{jia2004}
{Jia}, S.~M., {Chen}, Y., {Lu}, F.~J., {Chen}, L., \& {Xiang}, F. 2004, \aap,
  423, 65

\bibitem[Jimeno {et~al.}(2018)]{Jimeno2018}
Jimeno, P., Diego, J.~M., Broadhurst, T., {De Martino}, I., \& Lazkoz, R.
  2018, \mnras, 478, 638

\bibitem[{Kravtsov} \& {Borgani}(2012)]{kra2012}
{Kravtsov}, A.~V., \& {Borgani}, S. 2012, \araa, 50, 353

\bibitem[{Kravtsov} {et~al.}(2006)]{kra2006}
{Kravtsov}, A.~V., {Vikhlinin}, A., \& {Nagai}, D. 2006, \apj, 650, 128

\bibitem[{Lieu} {et~al.}(2006)]{lie2006}
{Lieu}, R., {Mittaz}, J.~P.~D., \& {Zhang}, S.-N. 2006, \apj, 648, 176

\bibitem[{Mahdavi} {et~al.}(2013)]{mah2013}
{Mahdavi}, A., {Hoekstra}, H., {Babul}, A., {et~al.} 2013, \apj, 767, 116

\bibitem[{Mantz} {et~al.}(2010)]{man2010i}
{Mantz}, A., {Allen}, S.~W., {Rapetti}, D., \& {Ebeling}, H. 2010, \mnras, 406,
  1759

\bibitem[{Maughan}(2007)]{Maughan2007}
{Maughan}, B.~J. 2007, \apj, 668, 772

\bibitem[{Mewe} {et~al.}(1985)]{mew1985}
{Mewe}, R., {Gronenschild}, E.~H.~B.~M., \& {van den Oord}, G.~H.~J. 1985,
  \aaps, 62, 197

\bibitem[{Morrison} \& {McCammon}(1983)]{mor1983}
{Morrison}, R., \& {McCammon}, D. 1983, \apj, 270, 119

\bibitem[{Motl} {et~al.}(2005)]{mot2005}
{Motl}, P.~M., {Hallman}, E.~J., {Burns}, J.~O., \& {Norman}, M.~L. 2005,
  \apjl, 623, L63

\bibitem[{Mroczkowski} {et~al.}(2009)]{mro2009}
{Mroczkowski}, T., {Bonamente}, M., {Carlstrom}, J.~E., {et~al.} 2009, \apj,
  694, 1034

\bibitem[Munari {et~al.}(2013)]{Munari2013}
Munari, E., Biviano, A., Borgani, S., Murante, G., \& Fabjan, D. 2013, \mnras,
  430, 2638

\bibitem[{Nagai}(2006)]{nag2006}
{Nagai}, D. 2006, \apj, 650, 538

\bibitem[{Planck Collaboration} {et~al.}(2011{\natexlab{a}})]{planck2011viii}
{Planck Collaboration}, {Ade}, P.~A.~R., {Aghanim}, N., {et~al.}
  2011{\natexlab{a}}, \aap, 536, A8

\bibitem[{Planck Collaboration} {et~al.}(2011{\natexlab{b}})]{planck2011x}
{Planck Collaboration}, {Aghanim}, N., {Arnaud}, M., {et~al.}
  2011{\natexlab{b}}, \aap, 536, A10

\bibitem[{Planck Collaboration} {et~al.}(2011{\natexlab{c}})]{planck2011xi}
{Planck Collaboration}, {Ade}, P.~A.~R., {Aghanim}, N., {et~al.}
  2011{\natexlab{c}}, \aap, 536, A11

\bibitem[{Planck Collaboration} {et~al.}(2013{\natexlab{a}})]{planck2011iii}
{Planck Collaboration}, {Ade}, P.~A.~R., {Aghanim}, N., {et~al.}
  2013{\natexlab{a}}, \aap, 550, A129

\bibitem[{Planck Collaboration} {et~al.}(2013{\natexlab{b}})]{planck2011v}
{Planck Collaboration}, {Ade}, P.~A.~R., {Aghanim}, N., {et~al.}
  2013{\natexlab{b}}, \aap, 550, A131

\bibitem[{Planck Collaboration} {et~al.}(2014)]{planck2013xxix}
{Planck Collaboration}, {Ade}, P.~A.~R., {Aghanim}, N., {et~al.} 2014, \aap,
  571, A29

\bibitem[{Planck Collaboration} {et~al.}(2016{\natexlab{a}})]{planck2015xxiv}
{Planck Collaboration}, {Ade}, P.~A.~R., {Aghanim}, N., {et~al.}
  2016{\natexlab{a}}, \aap, 594, A24

\bibitem[{Planck Collaboration} {et~al.}(2016{\natexlab{b}})]{planck2015xxvii}
{Planck Collaboration}, {Ade}, P.~A.~R., {Aghanim}, N., {et~al.}
  2016{\natexlab{b}}, \aap, 594, A27

\bibitem[Pratt {et~al.}(2009)]{Pratt2009}
Pratt, G.~W., Croston, J.~H., Arnaud, M., \& B{\"{o}}hringer, H. 2009, \aap,
  498, 361

\bibitem[Reichert {et~al.}(2011)]{Reichert2011}
Reichert, A., B{\"{o}}hringer, H., Fassbender, R., \& M{\"{u}}hlegger, M. 2011,
  \aap, 535, A4

\bibitem[{Reiprich} \& {B{\"o}hringer}(2002)]{rei2002}
{Reiprich}, T.~H., \& {B{\"o}hringer}, H. 2002, \apj, 567, 716

\bibitem[{Rozo} {et~al.}(2014{\natexlab{a}})]{roz2014ii}
{Rozo}, E., {Evrard}, A.~E., {Rykoff}, E.~S., \& {Bartlett}, J.~G.
  2014{\natexlab{a}}, \mnras, 438, 62

\bibitem[{Rozo} {et~al.}(2014{\natexlab{b}})]{roz2014i}
{Rozo}, E., {Rykoff}, E.~S., {Bartlett}, J.~G., \& {Evrard}, A.
  2014{\natexlab{b}}, \mnras, 438, 49

\bibitem[{Rozo} {et~al.}(2012)]{roz2012}
{Rozo}, E., {Vikhlinin}, A., \& {More}, S. 2012, \apj, 760, 67

\bibitem[{Rozo} {et~al.}(2010)]{roz2010}
{Rozo}, E., {Wechsler}, R.~H., {Rykoff}, E.~S., {et~al.} 2010, \apj, 708, 645

\bibitem[{Sayers} {et~al.}(2016)]{say2016}
{Sayers}, J., {Golwala}, S.~R., {Mantz}, A.~B., {et~al.} 2016, \apj, 832, 26

\bibitem[{Seljak} {et~al.}(2006)]{sel2006}
{Seljak}, U., {Slosar}, A., \& {McDonald}, P. 2006, \jcap, 10, 014

\bibitem[{Sembolini} {et~al.}(2014)]{sem2014iii}
{Sembolini}, F., {De Petris}, M., {Yepes}, G., {et~al.} 2014, \mnras, 440, 3520

\bibitem[Stanek {et~al.}(2006)]{Stanek2006}
Stanek, R., Evrard, A.~E., Bohringer, H., Schuecker, P., \& Nord, B. 2006,
  \apj, 648, 956

\bibitem[{Sunyaev} \& {Zeldovich}(1980)]{sun1980}
{Sunyaev}, R.~A., \& {Zeldovich}, I.~B. 1980, \araa, 18, 537

\bibitem[{Vikhlinin} {et~al.}(2009{\natexlab{a}})]{Vikhlinin2009}
{Vikhlinin}, A., {Burenin}, R.~A., {Ebeling}, H., {et~al.} 2009{\natexlab{a}},
  \apj, 692, 1033

\bibitem[{Vikhlinin} {et~al.}(2009{\natexlab{b}})]{Vikhlinin2009iii}
{Vikhlinin}, A., {Kravtsov}, A.~V., {Burenin}, R.~A., {et~al.}
  2009{\natexlab{b}}, \apj, 692, 1060

\bibitem[Zhang {et~al.}(2011)]{Zhang2011}
Zhang, Y.-Y., Andernach, H., Caretta, C.~A., {et~al.} 2011, \aap, 526, A105

\bibitem[{Zhao}(2015)]{zhathesis}
{Zhao}, H.-H. 2015, {The statistical studies on the X-ray properties of galaxy
  clusters}, PhD thesis, {Institute of High Energy Physics, Chinese Academy of
  Science, China}

\bibitem[{Zhao} {et~al.}(2013)]{zha2013}
{Zhao}, H.-H., {Jia}, S.-M., {Chen}, Y., {et~al.} 2013, \apj, 778, 124

\bibitem[{Zhao} {et~al.}(2015)]{zha2015}
{Zhao}, H.-H., {Li}, C.-K., {Chen}, Y., {Jia}, S.-M., \& {Song}, L.-M. 2015,
  \apj, 799, 47

\end{thebibliography}

\clearpage
\small
\begin{longtable}{crrrrccccc}
\caption{\label{tab:properties} Cluster properties}\\
\hline\hline
 name       &  RA & {Dec} & {z} & {$\theta_{500}$}  & \multicolumn{2}{c}{$\YP$}     &  $\YXMM$  & \multicolumn{2}{c}{Cool Core} \\
  & & & & & MaxSN  & NEAREST &   & Z13 & C07 \\
  &[deg]& [deg]  &         & [arcmin]         & [$\rm 10^{-4}arcmin^2$] & [$\rm 10^{-4}arcmin^2$] & [$\rm 10^{-4}arcmin^2$] & &\\
\hline
\endfirsthead
\caption{continued.}\\
\hline\hline
 name       &  RA & {Dec} & {z} & {$\theta_{500}$}  & \multicolumn{2}{c}{$\YP$}     &  $\YXMM$  & \multicolumn{2}{c}{Cool Core} \\
  & & & & & MaxSN  & NEAREST &   & Z13 & C07 \\
  &[deg]& [deg]  &         & [arcmin]         & [$\rm 10^{-4}arcmin^2$] & [$\rm 10^{-4}arcmin^2$] & [$\rm 10^{-4}arcmin^2$] & &\\
\hline
\endhead
\hline
\endfoot
2A0335          & 54.670 & 9.975 & 0.0347 & $22.4 \pm 0.2$ & $88.0 \pm 8.1$ & $91.0 \pm 10.5$ & $71.7 \pm 11.5$ & ${\surd}$ & ${\surd}$\\
A0085           & 10.459 & -9.305 & 0.0555 & $14.8 \pm 0.2$ & $88.4 \pm 5.0$ & $87.7 \pm 4.8$ & $90.7 \pm 7.7$ & ${\surd}$ & ${\surd}$\\
A0119           & 14.076 & -1.205 & 0.0444 & $11.6 \pm 0.8$ & $104.7 \pm 11.5$ & $104.7 \pm 11.5$ & $29.6 \pm 8.6$ & \texttimes & \texttimes\\
A0133           & 15.675 & -21.872 & 0.0569 & $16.2 \pm 1.5$ & $80.4 \pm 5.8$ & $44.5 \pm 3.9$ & $34.0 \pm 7.1$ & ${\surd}$ & ${\surd}$\\
A0399           & 44.457 & 13.049 & 0.0722 & $18.3 \pm 0.4$ & $135.1 \pm 10.2$ & $45.4 \pm 4.3$ & $84.3 \pm 27.2$ & \texttimes & \texttimes\\
A0401           & 44.740 & 13.579 & 0.0739 & $15.1 \pm 0.2$ & $90.0 \pm 6.8$ & $90.0 \pm 6.8$ & $113.6 \pm 12.3$ & \texttimes & \texttimes\\
A0478           & 63.356 & 10.467 & 0.0882 & $11.2 \pm 0.4$ & $75.1 \pm 5.1$ & $75.1 \pm 5.1$ & $111.9 \pm 7.4$ & ${\surd}$ & ${\surd}$\\
A0496           & 68.410 & -13.255 & 0.0326 & $23.0 \pm 0.8$ & $98.8 \pm 7.1$ & $87.7 \pm 6.5$ & $87.1 \pm 10.9$ & ${\surd}$ & ${\surd}$\\
A0576           & 110.343 & 55.786 & 0.0381 & $15.7 \pm 0.4$ & $46.7 \pm 5.4$ & $53.9 \pm 5.5$ & $28.5 \pm 5.8$ & ${\surd}$ & \texttimes\\
A0644           & 124.355 & -7.516 & 0.0704 & $15.5 \pm 0.6$ & $82.8 \pm 4.9$ & $77.9 \pm 4.0$ & $138.8 \pm 19.6$ & ${\surd}$ & ${\surd}$\\
A0754           & 137.285 & -9.655 & 0.0542 & $21.0 \pm 2.5$ & $213.2 \pm 15.2$ & $202.9 \pm 15.2$ & $89.8 \pm 21.2$ & \texttimes & \texttimes\\
A1413           & 178.827 & 23.407 & 0.1427 & $8.6 \pm 0.3$ & $28.2 \pm 2.1$ & $28.2 \pm 2.1$ & $36.4 \pm 4.4$ & \texttimes & ${\surd}$\\
A1644           & 194.291 & -17.405 & 0.0473 & $16.0 \pm 1.4$ & $54.0 \pm 5.8$ & $54.0 \pm 5.8$ & $73.4 \pm 12.0$ & ${\surd}$ & \texttimes\\
A1650           & 194.671 & -1.755 & 0.0845 & $10.7 \pm 0.1$ & $51.2 \pm 3.5$ & $51.2 \pm 3.5$ & $53.6 \pm 6.6$ & ${\surd}$ & \texttimes\\
A1651           & 194.840 & -4.188 & 0.0845 & $10.6 \pm 0.2$ & $45.2 \pm 3.5$ & $47.0 \pm 4.0$ & $57.8 \pm 5.7$ & \texttimes & ${\surd}$\\
A1689           & 197.875 & -1.338 & 0.1832 & $7.5 \pm 0.4$ & $40.2 \pm 2.2$ & $37.4 \pm 1.9$ & $34.3 \pm 4.4$ & \texttimes & ${\surd}$\\
A1775           & 205.474 & 26.372 & 0.0724 & $9.5 \pm 0.1$ & $16.5 \pm 2.6$ & $16.5 \pm 2.6$ & $18.7 \pm 1.6$ & ${\surd}$ & \texttimes\\
A1795           & 207.221 & 26.596 & 0.0622 & $19.3 \pm 0.1$ & $87.3 \pm 7.7$ & $87.3 \pm 7.7$ & $121.3 \pm 10.8$ & ${\surd}$ & ${\surd}$\\
A1914           & 216.507 & 37.827 & 0.1712 & $6.3 \pm 0.3$ & $38.7 \pm 3.3$ & $26.0 \pm 1.6$ & $35.8 \pm 3.0$ & \texttimes & ${\surd}$\\
A2029           & 227.729 & 5.720 & 0.0766 & $11.5 \pm 0.5$ & $91.5 \pm 11.8$ & $111.0 \pm 11.8$ & $119.3 \pm 10.8$ & ${\surd}$ & ${\surd}$\\
A2063           & 230.772 & 8.602 & 0.0358 & $15.7 \pm 0.5$ & $38.7 \pm 5.2$ & $38.7 \pm 5.2$ & $27.4 \pm 6.5$ & \texttimes & ${\surd}$\\
A2065           & 230.611 & 27.709 & 0.0723 & $11.0 \pm 0.2$ & $42.6 \pm 3.1$ & $65.5 \pm 4.8$ & $50.2 \pm 6.7$ & ${\surd}$ & \texttimes\\
A2142           & 239.586 & 27.227 & 0.0894 & $11.0 \pm 0.5$ & $117.2 \pm 15.1$ & $156.9 \pm 15.1$ & $93.8 \pm 20.2$ & ${\surd}$ & ${\surd}$\\
A2163           & 243.945 & -6.138 & 0.2030 & $12.2 \pm 0.3$ & $145.5 \pm 9.1$ & $145.5 \pm 9.1$ & $199.4 \pm 40.4$ & \texttimes & \texttimes\\
A2199           & 247.158 & 39.549 & 0.0299 & $22.8 \pm 0.3$ & $110.8 \pm 7.5$ & $114.3 \pm 6.6$ & $124.6 \pm 51.4$ & ${\surd}$ & ${\surd}$\\
A2204           & 248.194 & 5.571 & 0.1514 & $8.6 \pm 0.3$ & $44.9 \pm 2.8$ & $41.7 \pm 2.9$ & $62.3 \pm 10.3$ & ${\surd}$ & ${\surd}$\\
A2255           & 258.197 & 64.061 & 0.0809 & $10.8 \pm 1.0$ & $74.7 \pm 8.5$ & $74.7 \pm 8.5$ & $41.0 \pm 10.7$ & \texttimes & \texttimes\\
A2256           & 255.953 & 78.644 & 0.0581 & $13.8 \pm 0.4$ & $111.8 \pm 8.5$ & $111.8 \pm 8.5$ & $74.0 \pm 10.3$ & \texttimes & \texttimes\\
A2319           & 290.298 & 43.948 & 0.0564 & $21.3 \pm 0.2$ & $273.5 \pm 32.9$ & $247.6 \pm 32.9$ & $88.0 \pm 32.7$ & \texttimes & \texttimes\\
A2589           & 350.987 & 16.775 & 0.0416 & $16.0 \pm 0.4$ & $57.0 \pm 12.2$ & $31.6 \pm 5.7$ & $38.0 \pm 3.6$ & \texttimes & ${\surd}$\\
A2597           & 351.333 & -12.122 & 0.0852 & $7.5 \pm 0.1$ & $11.3 \pm 2.4$ & $9.8 \pm 2.3$ & $8.5 \pm 2.9$ & ${\surd}$ & ${\surd}$\\
A2657           & 356.238 & 9.198 & 0.0400 & $17.3 \pm 1.1$ & $33.6 \pm 5.0$ & $33.6 \pm 5.0$ & $31.5 \pm 10.4$ & ${\surd}$ & ${\surd}$\\
A2734           & 2.836 & -28.855 & 0.0620 & $11.4 \pm 0.6$ & $42.6 \pm 4.2$ & $42.6 \pm 4.2$ & $26.4 \pm 6.0$ & ${\surd}$ & \texttimes\\
A3112           & 49.494 & -44.238 & 0.0752 & $11.8 \pm 0.2$ & $28.7 \pm 3.2$ & $35.2 \pm 3.4$ & $39.2 \pm 2.6$ & ${\surd}$ & ${\surd}$\\
A3158           & 55.725 & -53.638 & 0.0590 & $12.6 \pm 0.4$ & $50.0 \pm 12.2$ & $50.0 \pm 12.2$ & $47.7 \pm 5.0$ & \texttimes & \texttimes\\
A3266           & 67.850 & -61.438 & 0.0589 & $19.6 \pm 1.0$ & $173.5 \pm 16.7$ & $161.1 \pm 16.7$ & $199.1 \pm 51.4$ & \texttimes & \texttimes\\
A3391           & 96.595 & -53.688 & 0.0514 & $17.9 \pm 0.7$ & $48.9 \pm 6.3$ & $48.9 \pm 6.3$ & $48.5 \pm 16.5$ & \texttimes & \texttimes\\
A3526           & 192.200 & -41.305 & 0.0114 & $54.9 \pm 0.7$ & $211.8 \pm 33.0$ & $211.8 \pm 33.0$ & $287.9 \pm 21.1$ & ${\surd}$ & ${\surd}$\\
A3532           & 194.320 & -30.372 & 0.0554 & $11.5 \pm 0.9$ & $68.5 \pm 5.2$ & $87.4 \pm 6.3$ & $35.9 \pm 8.2$ & \texttimes & \texttimes\\
A3558           & 201.990 & -31.505 & 0.0488 & $14.5 \pm 0.9$ & $36.9 \pm 2.3$ & $36.9 \pm 2.3$ & $113.9 \pm 17.9$ & ${\surd}$ & \texttimes\\
A3562           & 203.401 & -31.655 & 0.0490 & $14.6 \pm 0.3$ & $159.8 \pm 21.0$ & $98.5 \pm 12.3$ & $32.6 \pm 3.1$ & \texttimes & \texttimes\\
A3571           & 206.868 & -32.838 & 0.0391 & $22.4 \pm 0.6$ & $163.0 \pm 9.1$ & $163.0 \pm 9.1$ & $236.0 \pm 22.0$ & \texttimes & ${\surd}$\\
A3667           & 303.127 & -56.822 & 0.0556 & $18.0 \pm 0.3$ & $178.3 \pm 19.8$ & $178.3 \pm 19.8$ & $246.5 \pm 8.8$ & \texttimes & \texttimes\\
A3695           & 308.700 & -35.805 & 0.0894 & $9.3 \pm 0.4$ & $19.3 \pm 3.6$ & $30.0 \pm 3.9$ & $27.4 \pm 5.1$ & \texttimes & \texttimes\\
A3822           & 328.538 & -57.855 & 0.0760 & $8.2 \pm 0.7$ & $11.1 \pm 2.9$ & $11.1 \pm 2.9$ & $17.3 \pm 3.8$ & \texttimes & \texttimes\\
A3827           & 330.483 & -59.938 & 0.0980 & $10.2 \pm 0.2$ & $48.3 \pm 2.5$ & $48.3 \pm 2.5$ & $51.8 \pm 5.0$ & \texttimes & \texttimes\\
A3888           & 338.629 & -37.738 & 0.1510 & $6.3 \pm 0.8$ & $47.7 \pm 4.0$ & $27.5 \pm 1.9$ & $27.6 \pm 3.2$ & \texttimes & \texttimes\\
A4038           & 356.930 & -28.138 & 0.0300 & $19.7 \pm 0.3$ & $42.2 \pm 4.8$ & $42.2 \pm 4.8$ & $47.9 \pm 2.9$ & ${\surd}$ & ${\surd}$\\
A4059           & 359.260 & -34.755 & 0.0475 & $14.7 \pm 0.2$ & $70.7 \pm 6.0$ & $70.7 \pm 6.0$ & $59.3 \pm 9.2$ & ${\surd}$ & ${\surd}$\\
AWM7            & 43.623 & 41.578 & 0.0172 & $36.7 \pm 1.4$ & $202.6 \pm 12.8$ & $202.6 \pm 12.8$ & $153.8 \pm 52.1$ & ${\surd}$ & ${\surd}$\\
Coma            & 194.929 & 27.939 & 0.0231 & $51.8 \pm 2.1$ & $1019.5 \pm 43.7$ & $1519.1 \pm 43.7$ & $1210. \pm 440.$ & \texttimes & \texttimes\\
MKW3s           & 230.458 & 7.709 & 0.0442 & $16.3 \pm 1.0$ & $30.9 \pm 5.5$ & $50.1 \pm 13.0$ & $34.0 \pm 4.7$ & ${\surd}$ & ${\surd}$\\
RXCJ2344.2-0422 & 356.067 & -4.372 & 0.0786 & $8.5 \pm 0.3$ & $27.1 \pm 3.1$ & $20.4 \pm 2.9$ & $21.2 \pm 4.1$ & \texttimes & \texttimes\\
S0636           & 157.515 & -35.309 & 0.0116 & $30.9 \pm 2.1$ & $84.2 \pm 12.1$ & $84.2 \pm 12.1$ & $25.5 \pm 12.9$ & ${\surd}$ & \texttimes\\
Triangulum      & 249.576 & -64.356 & 0.0510 & $21.2 \pm 0.8$ & $244.1 \pm 50.3$ & $244.1 \pm 50.3$ & $387.0 \pm 68.7$ & \texttimes & \texttimes\\
\hline
A1835           & 210.260 & 2.880 & 0.2528 & $5.2 \pm 0.3$ & $23.4 \pm 1.5$ & $23.4 \pm 1.5$ & $23.5 \pm 5.2$ & ${\surd}$ &   -   \\
A2034           & 227.549 & 33.515 & 0.1130 & $7.9 \pm 0.3$ & $37.8 \pm 4.3$ & $30.8 \pm 2.0$ & $24.3 \pm 3.2$ & \texttimes &   -   \\
A2219           & 250.089 & 46.706 & 0.2280 & $5.9 \pm 0.2$ & $42.2 \pm 5.3$ & $42.2 \pm 5.3$ & $45.0 \pm 6.7$ & \texttimes &   -   \\
A2390           & 328.398 & 17.687 & 0.2329 & $6.5 \pm 0.3$ & $33.2 \pm 1.9$ & $47.9 \pm 3.8$ & $64.8 \pm 6.6$ & ${\surd}$ &   -   \\
A2420           & 332.582 & -12.172 & 0.0846 & $13.7 \pm 0.4$ & $50.6 \pm 3.3$ & $47.5 \pm 3.3$ & $58.9 \pm 12.3$ & \texttimes &   -   \\
A2426           & 333.636 & -10.372 & 0.0980 & $9.7 \pm 0.3$ & $40.8 \pm 5.0$ & $25.8 \pm 3.1$ & $13.5 \pm 2.2$ & \texttimes &   -   \\
A2626           & 354.126 & 21.142 & 0.0565 & $12.9 \pm 0.5$ & $48.4 \pm 19.2$ & $48.4 \pm 19.2$ & $15.2 \pm 1.8$ & ${\surd}$ &   -   \\
A3186           & 58.095 & -74.014 & 0.1279 & $7.1 \pm 0.6$ & $30.5 \pm 5.3$ & $30.5 \pm 5.3$ & $34.1 \pm 7.7$ & \texttimes &   -   \\
A3404           & 101.372 & -54.222 & 0.1644 & $10.6 \pm 0.3$ & $53.5 \pm 5.6$ & $53.5 \pm 5.6$ & $60.0 \pm 26.1$ & ${\surd}$ &   -   \\
A3911           & 341.577 & -52.722 & 0.0965 & $11.4 \pm 0.7$ & $34.8 \pm 2.8$ & $34.8 \pm 2.8$ & $33.9 \pm 7.7$ & \texttimes &   -   \\
RXCJ0413.9-3805 & 63.488 & -38.088 & 0.0501 & $14.3 \pm 0.2$ & $22.3 \pm 3.8$ & $22.3 \pm 3.8$ & $9.5 \pm 2.1$ & \texttimes &   -   \\
RXCJ1504.1-0248 & 226.032 & -2.805 & 0.2153 & $5.2 \pm 0.1$ & $12.4 \pm 2.3$ & $11.9 \pm 2.2$ & $18.7 \pm 3.8$ & ${\surd}$ &   -   \\
RXCJ1558.3-1410 & 239.597 & -14.172 & 0.0970 & $7.6 \pm 0.3$ & $15.5 \pm 3.4$ & $15.5 \pm 3.4$ & $12.7 \pm 1.7$ & ${\surd}$ &   -   \\
RXCJ1720.1+2637 & 260.039 & 26.627 & 0.1644 & $6.9 \pm 0.2$ & $22.4 \pm 2.3$ & $19.1 \pm 2.0$ & $28.2 \pm 3.9$ & ${\surd}$ &   -   \\
RXCJ2014.8-2430 & 303.707 & -24.505 & 0.1612 & $6.5 \pm 0.4$ & $12.1 \pm 2.0$ & $12.1 \pm 2.0$ & $23.3 \pm 6.4$ & ${\surd}$ &   -   \\

\end{longtable}

\end{document}